\documentclass[prd,showpacs,amsmath,twocolumn,amssymb,superscriptaddress,nofootinbib,showkeys]{revtex4}
\usepackage{amssymb}
\usepackage{amsmath}

\newcommand{\Fst}{{\mathop {\rule{0pt}{0pt}{F}}\limits^{\;*}}\rule{0pt}{0pt}}

\usepackage[unicode]{hyperref}

\begin{document}
\title{Axion-induced oscillations of cooperative electric field \\ in a cosmic magneto-active plasma}
\author{Alexander B. Balakin}
\email{Alexander.Balakin@kpfu.ru} \affiliation{Department of
General Relativity and Gravitation, Institute of Physics, Kazan
Federal University, Kremlevskaya str. 18, Kazan 420008, Russia}

\author{Ruslan K. Muharlyamov}
\email{Ruslan.Muharlyamov@kpfu.ru} \affiliation{Department of
General Relativity and Gravitation, Institute of Physics, Kazan
Federal University, Kremlevskaya str. 18, Kazan 420008, Russia}

\author{Alexei E. Zayats}
\email{Alexei.Zayats@kpfu.ru} \affiliation{Department of General
Relativity and Gravitation, Institute of Physics, Kazan Federal
University, Kremlevskaya str. 18, Kazan 420008, Russia}


\begin{abstract}
We consider one cosmological application of an axionic extension
of the Maxwell-Vlasov theory, which describes axionically induced
oscillatory regime in the state of global magnetic field evolving
in the anisotropic expanding (early) universe. We show that the
cooperative electric field in the relativistic plasma, being
coupled to the pseudoscalar (axion) and global magnetic fields,
plays the role of a regulator in this three-level system; in
particular, the cooperative (Vlasov) electric field converts the
regime of anomalous growth of the pseudoscalar field, caused by
the axion-photon coupling at the inflationary epoch of the
universe expansion, into an oscillatory regime with finite density
of relic axions. We analyze solutions to the dispersion equations
for the axionically induced cooperative oscillations of the
electric field in the relativistic plasma.

\end{abstract}



\maketitle

\section{Introduction}

The Vlasov concept of a cooperative macroscopic electromagnetic
field in a collisionless plasma \cite{Vlasov1,Vlasov2} (or,
equivalently, self-consistent, collective, average, mean field) is
nowadays one of corner-stones of modern plasma theory (see, e.g.,
\cite{MV1,MV2,Silin,2}). The Maxwell-Vlasov theory, based on this
concept, deals in fact with a two-level self-regulating system:
electrically charged plasma particles move under the influence of
a macroscopic electromagnetic field and produce the cooperative
electric current, which, in its turn, generates the
self-consistent electromagnetic field. Thus, the ensemble of
plasma particles controls itself by means of cooperative
electromagnetic field. Master equations of the Maxwell-Vlasov
theory include, respectively, two coupled sub-systems:
electrodynamic and kinetic equations.

When we consider the Maxwell-Vlasov-axion model, we deal with a
self-consistent theory of interaction between electromagnetic,
pseudoscalar (axion) fields and a multi-component plasma. This
physical system can be indicated as the three-level one, since in
addition to plasma particles and photons it includes axions, i.e.,
hypothetical light massive pseudo-bosons, which appeared in the
lexicon of High Energy Physics in the context of a strong
CP-violation problem and spontaneous breaking of symmetry (see,
e.g., \cite{Peccei,Weinberg,Wilczek0,Peccei2,Battesti}), and are
considered as dark matter candidates (see, e.g.,
\cite{Raffelt,Turner,Khl}). The contribution of the pseudoscalar
field into the electrodynamic equations was discussed, first, by
Ni \cite{8Ni}; the axion electrodynamics as an accomplished
science appeared in the eighties of last century (see, e.g.,
\cite{Sikivie,Wil}). The corresponding master equations of the
Maxwell-Vlasov-axion model includes, first, equations of the axion
electrodynamics (instead of the Faraday-Maxwell equations),
second, the equation for the evolution of the pseudoscalar (axion)
field, third, the kinetic equation for the plasma. This
Maxwell-Vlasov-axion model is the corresponding part of the
Einstein-Maxwell-Vlasov-axion model elaborated by authors in
\cite{BMZ1} and applied to cosmology in \cite{BMZ2}.

The axionic extension of Vlasov's idea prompted us to consider a
cooperative pseudoscalar (axion) field, and to analyze Vlasov-type
models of the axion-active plasma (we introduced this term in
analogy with the magneto-active plasma). ``Modus operandi'' of the
new axion cooperative field in plasma seems to be the following:
axion field regulates the electromagnetic one via specific
current-type term in the electrodynamic equations and (probably)
form the axionic force in the kinetic equation; in its turn, the
state of pseudoscalar (axion) field is predetermined by two
pseudoscalar sources: of the electromagnetic origin and induced by
plasma particles (see \cite{BMZ1} for details).

We expect that this axionically extended model can be interesting
for applications to cosmology and astrophysics. Indeed, the relic
axions are assumed to form the cold dark matter, the contribution
of which into the universe energy balance is estimated to be about
$23\%$  \cite{DM1,DM2,DM3}. The mass density of the dark matter in
the Solar system is estimated to be $\rho_{({\rm DM})} \simeq
0.033 \ M_{({\rm Sun})} {\rm pc}^{-3}$ or in the natural units
$\rho_{({\rm DM})} \simeq 1.25 \ {\rm GeV} \cdot {\rm cm}^{-3}$.
Taking into account that the axion mass is assumed to belong to
the interval $10^{-6}\ {\rm eV} \lesssim m_{(\rm axion)}\lesssim
10^{-2}\ {\rm eV}$, the axion number density in the Earth vicinity
can be (optimistically) estimated as $N_{({\rm A})} \simeq 10^{11}
{-} 10^{15}\ {\rm cm}^{-3}$. In addition, plasma is an important
constituent of many objects and media in our universe, and the
photons emitted, scattered or deflected by the plasma particles,
propagate in the environment of axionic dark matter. Thus, models
of the axion-photon coupling in a relativistic plasma can clarify
some properties of the electromagnetic waves emitted by
astrophysical sources and detected by astronomers.

Axion-photon coupling is nowadays studied  by many experimental
groups; the most recent data concerning the modern status of these
experiments can be found, e.g., in the reports of Collaborations
abbreviated as PVLAS, GammeV, CAST, OSQAR, Q\&A, BMV, ADMX, ALPS,
XENON, EDELWEISS, CDMS, etc., published in the proceedings of last
Patras Workshop on Axions and WIMPs \cite{Patras}.

In this paper we consider one application of the model established
in \cite{BMZ1}: the cooperative oscillations in the relativistic
plasma, coupled to the axionic dark matter and to the global
magnetic field, in the framework of the Bianchi-I cosmological
model, the analysis of dispersion relations being the main purpose
of the work.

\section{The model}

\subsection{Kinetic equation}

The first key element of the Maxwell-Vlasov-axion model is the relativistic kinetic equation
\begin{equation}
g^{is}p_s\left[\frac{\partial}{\partial x^i}+(\Gamma_{ik}^l
p_l+e_{({\rm a})}F_{ki})\frac{\partial}{\partial
p_k}\right]f_{({\rm a})}=0 \,. \label{kin1}
\end{equation}
Here $f_{({\rm a})}(x^i, p_k)$ is the 8-dimensional one-particle
distribution function, which describes particles of a sort $({\rm
a})$ with the rest mass $m_{({\rm a})}$ and electric charge
$e_{({\rm a})}$; this function depends on four coordinates $x^i$
and on the momentum four-covector $p_k$. The term $e_{({\rm
a})}F_{k}^{\ i}p_i$ introduces the Lorentz force; $F_{ik}$ is the
Maxwell tensor, $\Gamma^l_{ik}$ are the Christoffel symbols
associated with the spacetime metric $g_{ik}$. Characteristic
equations associated with the kinetic equation (\ref{kin1}) are
the following:
\begin{gather}
\frac{dp_k}{ds}- (\Gamma_{ik}^l p_l +e_{({\rm
a})}F_{ki})\frac{dx^i}{ds} =0, {}\nonumber\\ \frac{dx^i}{ds} =
\frac{1}{m_{({\rm a})}} g^{ij} p_j \,. \label{kin3}
\end{gather}
The four-interval $ds$ is in this context a proper time for plasma particles.

\subsection{Electrodynamic equations}

The second key element of the model is the axionically extended Maxwell equations
\begin{equation}
\nabla_k  F^{ik} = - \Fst^{ik}\nabla_k \phi - I^i \,. \label{eld1}
\end{equation}
The term $\Fst^{ik} \equiv \frac{1}{2} \epsilon^{ikmn}F_{mn}$
describes the tensor dual to the Maxwell tensor $F_{mn}$;
$\epsilon^{ikmn} \equiv \frac{1}{\sqrt{-g}} E^{ikmn}$ is the
Levi-Civita tensor, $E^{ikmn}$ is the absolutely antisymmetric
Levi-Civita symbol with $E^{0123}{=}1$. We use the
dimensionless pseudoscalar field $\phi$ for description of axions.
For such approach the axion-photon coupling constant appears in
front of the kinetic term of the axion field Lagrangian (see,
e.g., \cite{BMZ1} for details). Rescaling of this type is
convenient for several problem of axion electrodynamics
\cite{BBT,BG}, since two principal terms in the Lagrangian, $\frac14 F_{mn}F^{mn}$ and $\frac14 \phi F^*_{mn}F^{mn}$, have explicitly
the same dimensionality.

The dual Maxwell tensor satisfies the condition
\begin{equation}
\nabla_{k} \Fst^{ik} =0 \,, \label{Emaxstar}
\end{equation}
$\nabla_{k}$ is the covariant derivative.
The electric current four-vector $I^i$ is considered to consist in
a linear combination of first moments of the distribution
functions $f_{({\rm a})}(x^l, p_k)$
\begin{gather}
I^i = \sum_{({\rm a})} \frac{e_{({\rm a})}}{m_{({\rm a})}}\int
\frac{d_4P}{\sqrt{-g}}g^{ij}p_j f_{({\rm a})}(x^l, p_k) \nonumber\\
{}\times\delta(\sqrt{g^{sn}p_s p_n} -m_{({\rm a})})\Theta(V^h p_h)
\,, \label{kin01}
\end{gather}
where $d_4P=dp_0dp_1dp_2dp_3$ symbolizes the volume in the
four-dimensional momentum space; the delta function guarantees the
normalization property of the particle momentum, $g^{kl}p_k p_l
=m^2$; the Heaviside function $\Theta(V^h p_h)$ rejects negative
values of energy, $V^h$ is the velocity four-vector of the system
as a whole; $g$ is the determinant of the tensor $g_{ik}.$

\subsection{Equation for the pseudoscalar (axion) field}

The third element of the model is the master equation for the
pseudoscalar field $\phi$:
\begin{gather}
 \left[\nabla_m\nabla^m + m^2_{({\rm A })}\right] \phi = \frac{1}{4\Psi^2_0} \Fst_{mn}F^{mn}\,.
\label{ax1}
\end{gather}
Here $\phi$ is a dimensionless quantity. The parameter $m_{({\rm A
})}$ is defined as $m_{({\rm A })}{=}\frac{m_{(\rm axion)}
c}{\hbar}$, where $m_{(\rm axion)}$ is the axion mass (let us
remind that we use the natural units with $\hbar{=}1$, $c{=}1$).
Estimations of the axion mass give the interval $10^{{-}6}\ {\rm
eV} \lesssim m_{(\rm axion)}\lesssim 10^{{-}2}\ {\rm eV}$. The
parameter $\frac{1}{\Psi_0}$ is the coupling constant of the
axion-photon interaction; its estimations give $10^{5}\ {\rm GeV}
\lesssim \Psi_0 \lesssim 10^{12}\ {\rm GeV}$ (see, e.g.,
\cite{Patras}).

\subsection{Gravity field description}

In the Maxwell-Vlasov-axion model we consider the gravitational field to be given and the space-time background to be described by the metric
\begin{equation}
ds^2 = dt^2-a^2(t)\left[(dx^1)^2+(dx^2)^2\right]-c^2(t)(dx^3)^2
\,, \label{1gr}
\end{equation}
which relates to the well-known Bianchi-I anisotropic homogeneous
cosmological model with local rotational isotropy. We consider the
scale factors $a(t)$ and $c(t)$ to be known functions of
cosmological time $t$. Also, we assume that the pseudoscalar,
electric and magnetic fields inherit the spacetime symmetry: they
depend on time only, as well as, the electric and magnetic fields
are directed along the anisotropy axis, i.e., $B^3(t)\neq 0$ and
$E^3(t)\neq 0$ only. Let us stress that in this work we do not
consider the electromagnetic and axion perturbations depending
(generally) on all four coordinates; we study the oscillations
only, i.e., we restrict ourselves by the solutions depending on
time only. General analysis of the dispersion relations in the
axion-magneto-active plasma is planned to be fulfilled in a
separate paper.

\section{Solutions to the master equations in the framework of Bianchi-I model}

\subsection{Solution to the Vlasov equation}

The Vlasov equation (\ref{kin1}) is the homogeneous differential
equation of the first order in partial derivatives, thus, the
distribution function $f_{({\rm a})}$ is arbitrary function of
seven integrals of motion. One of the integrals, is clearly, the
quadratic quantity $C_0 = g^{ik}p_ip_k$. For other integrals we
have to fix the structure of the Maxwell tensor $F_{ik}$; we
assume that it contains two non-vanishing components only:
\begin{equation}
F_{12} = - a^2(t)c(t) B^{3} \,, \quad F_{30} = - c^2(t)E^3 =
-\frac{d A_3 }{d t}   \,. \label{Maxwell1}
\end{equation}
Here $A_3(t)$ is the longitudinal component of the electromagnetic potential four-vector.
As for the component $F_{12}$, it is clear from (\ref{Emaxstar}) that it is a constant:
\begin{equation}
\nabla_k F^{*ik} = 0  \ \rightarrow \ \frac{d}{dt}[E^{i012}F_{12}]
= 0  \ \rightarrow \  F_{12} = {\rm const}\,. \label{Maxwell11}
\end{equation}
Keeping in mind the characteristic equation $m_{({\rm a})}
\frac{dt}{ds}=p_0$, we readily obtain the so-called longitudinal
integral of motion as follows:
\begin{equation}
dp_3 = e_{({\rm a})}F_{30} dt  \ \ \rightarrow \ \ C_{||} = p_3 +
e_{({\rm a})} A_3 \,. \label{long}
\end{equation}
Two transversal integrals of motion can be found from the characteristic equations
\begin{equation}
\frac{dp_2}{p_1} = - \frac{dp_1}{p_2} =  \frac{e_{({\rm a})}F_{12}
\ dt}{p_0(t) a^2(t)}  \,. \label{tr1}
\end{equation}
The solution of (\ref{tr1}) is known to have the form
\begin{gather}
p_1 = C_{\bot} \cos{\Phi(t)} \,, \quad p_2 = C_{\bot}
\sin{\Phi(t)} \,, \label{tr2} \\
\Phi(t) = \Phi(0) + e_{({\rm
a})}F_{12} \int\limits_0^t \frac{d\tau}{p_0(\tau)a^2(\tau)} \,.
\label{tr3}
\end{gather}
Clearly, we recover two well-known integrals of motion
\begin{gather}
C^2_{\bot} =p^2_{\bot} \equiv p_1^2+p_2^2   \,,\nonumber\\
\Phi(0) = \arctan{\frac{p_2}{p_1}} - e_{({\rm a})}F_{12}
\int\limits_0^t \frac{d\tau}{p_0(\tau)a^2(\tau)} \,. \label{trans}
\end{gather}
The component $p_0$ of the particle momentum four-vector, which we
need to calculate $\Phi(t)$, can be found from the quadratic
integral of motion $g^{ij}p_ip_j=m^2_{({\rm a})}=C_0$:
\begin{equation}
p_{0}(t)=p^0(t)=\sqrt{m^2_{({\rm
a})}+\frac{C^2_{\bot}}{a^2(t)}+\frac{[C_{||}-e_{({\rm
a})}A_3(t)]^2}{c^2(t)}} \,. \label{kin5}
\end{equation}
Other three integrals of motion are
\begin{gather}
C_4 =x^1(t)+\frac{C_{\bot}}{e_{({\rm a})}F_{12}} \sin{\Phi(t)}
\,,\nonumber\\
C_5=x^2(t)-\frac{C_{\bot}}{e_{({\rm a})}F_{12}} \cos{\Phi(t)}
\,,\nonumber\\
C_6 = x^3(t)+\int\limits_0^t [C_{||}-e_{({\rm
a})}A_3(\tau)]\frac{d \tau}{p^0(\tau)c^2(\tau)} \,. \label{int99}
\end{gather}
Generally, the distribution function, which satisfies kinetic
equation (\ref{kin1}) can be reconstructed as an arbitrary
function of seven integrals of motion: $C_0$, $C_{\bot}$,
$\Phi(0)$, $C_{||}$, $C_4$, $C_5$ and $C_6$. Taking into account
that the spacetime is anisotropic and homogeneous, we require that
all the macroscopic moments of the distribution function
\begin{equation}
T_{i_1...i_s}(x) \equiv \int \frac{d_4P}{\sqrt{-g}} f(x^i,p_k)
p_{i_1}...p_{i_s} \label{Lie1}
\end{equation}
inherit this symmetry, i.e., the Lie derivatives of the tensors $T_{i_1...i_s}(x)$ (with arbitrary $s$)
calculated along the Killing vectors $\xi_{\alpha}$ are equal to zero \cite{Yano}:
\begin{align}
\pounds_{\xi_{\alpha}}T_{i_1...i_s} &= \xi^l_{\alpha} \partial_l
T_{i_1...i_s} \nonumber \\{} &+ T_{l...i_s}\partial_{i_1}
\xi^{l}_{\alpha} +\ldots +T^{i_1...l}\partial_{i_s}
\xi^{l}_{\alpha} = 0 \,. \label{Lie2}
\end{align}
As was shown in \cite{Ivanov}, the relationships (\ref{Lie2}) are satisfied for arbitrary $s$, when the distribution function satisfies the following equations:
\begin{equation}
    \xi^k_\alpha \frac{\partial f}{\partial x^k}+ (\partial_k\xi^i_\alpha) \ p^k \frac{\partial f}{\partial p^i}=0 \,.
\label{Lie3}
\end{equation}
The parameter $\alpha$ indicates the Killing vectors, which describe the spacetime symmetry; in the Bianchi-I model with local rotational symmetry we deal with four Killing vectors
\begin{equation}
    \xi^k_{1}{=}\delta_1^k \,,\quad \xi^k_{2}{=}\delta_2^k \,,\quad \xi^k_{3}{=}\delta_3^k \,, \quad
    \xi^k_{4}{=}x^1\delta^k_2{-}x^2\delta^k_1 \,.
    \label{Lie4}
\end{equation}
For first three Killing vectors the relationships (\ref{Lie3}) yield
\begin{equation}
    \frac{\partial f}{\partial x^1}=\frac{\partial f}{\partial x^2}=\frac{\partial f}{\partial x^3} = 0 \,,
    \label{Lie5}
    \end{equation}
i.e., the distribution function does not depend on spatial coordinates $x^1$, $x^2$ and $x^3$.
The fourth Killing vector provides the condition
\begin{equation}
p^1\,\frac{\partial f}{\partial p^2}-p^2\,\frac{\partial f}{\partial p^1}=0 \,,
\label{Lie6}
\end{equation}
which relates to the local rotational symmetry of the model.
The conditions (\ref{Lie5}) require that the integrals $C_4$, $C_5$, $C_6$ (see (\ref{int99})) can not appear as the arguments of the distribution function.
The condition  (\ref{Lie6}) holds, when $p_1$  and $p_2$ enter the distribution function only as $p_2^2{+}p_3^2$. Thus, the distribution function has to include
only three integrals of motion from seven, and should have the form
\begin{equation}
f_{({\rm a})}(x^l, p_k) = f^{(0)}_{({\rm a})}(C^2_{\bot},
C^2_{||}) \delta(\sqrt{C_0}-m_{({\rm a})}) \,. \label{kin7}
\end{equation}
Here $f^{(0)}_{({\rm a})}(C^2_{\bot}, C^2_{||})$ is arbitrary
function of two arguments. As for the the macroscopic velocity
four-vector $V^h$, as usual for the Bianchi-I type spacetime, we
put $V^h=\delta^h_0$. Then the cooperative electric current $I^i$
happens to be reduced to the following term: $I^i =  \delta^i_3
\cdot I^3$
\begin{gather}
I^3 =-\sum_{({\rm a})} \frac{\pi e_{({\rm
a})}}{a^2(t)c^3(t)}\int\limits_ 0^{\infty} d C^2_{\bot}
\int\limits_{-\infty}^{\infty}dC_{||}
f^{(0)}_{({\rm a})}(C^2_{\bot}, C^2_{||}) \nonumber\\
\times \frac{[C_{||}{-}e_{({\rm a})}A_3(t)]}{\sqrt{m^2_{({\rm
a})}+\frac{C^2_{\bot}}{a^2(t)}+\frac{[C_{||}-e_{({\rm
a})}A_3]^2}{c^2(t)}}} \,. \label{kin8}
\end{gather}
Clearly, $I^1=I^{2}=0$ since $f^{(0)}_{({\rm a})}$ is even
function of $p_1$ and $p_2$. The component $I^0$ vanishes since
the plasma is considered to be electro-neutral. The component
$I^3$ as a function of the potential $A_3(t)$ displays the
following properties:
\begin{equation}
I^3 \left(A_3 \to 0 \right) \to 0 \,, \quad I^3 \left(A_3 \to \infty \right) \to 0 \,.
\label{kin87}
\end{equation}
Finally, when the energy $e_{({\rm a})}A_3$ of the plasma particle
in the electric field is much smaller than its rest mass, i.e.,
$|e_{({\rm a})}A_3|\ll m_{({\rm a})}$, one can simplify the
current as follows:
\begin{equation}
c^2(t)I^3(A_3)  \to  \Omega^2_{\rm L} \cdot A_3(t) \,,
\label{kin81}
\end{equation}
where the term
\begin{gather}
\Omega^2_{\rm L} \equiv \sum_{({\rm a})} \frac{\pi e^2_{({\rm
a})}}{a^2 c}\int\limits_0^{\infty} d C^2_{\bot}
\int\limits_{-\infty}^{\infty}dC_{||} f^{(0)}_{a}(C^2_{\bot},
C^2_{||}) \nonumber\\
{}\times \frac{\left[m^2_{({\rm
a})}+\frac{C^2_{\bot}}{a^2(t)}\right]}{\left[m^2_{({\rm
a})}+\frac{C^2_{\bot}}{a^2(t)}+\frac{C^2_{||}}{c^2(t)}\right]^{3/2}}
\,. \label{kin83}
\end{gather}
is a generalization of the well-known Langmuir frequency
\cite{Silin,MV1} (indeed, in the non-relativistic limit
$\Omega^2_{\rm L}=\sum\limits_{({\rm a})} \frac{4\pi e^2_{({\rm
a})} {\cal N}_{({\rm a})}}{m_{({\rm a})}}$, where ${\cal N}_{({\rm
a})}$ is the particle number density).

\subsection{Solutions to the equations of axion electrodynamics}

The Maxwell equations (\ref{eld1}) reduce now to one linear differential equation for the function $A_3(t)$:
\begin{equation}
{\ddot{A}}_3 + {\dot{A}}_3 \left(\frac{2\dot a}{a}-\frac{\dot
c}{c}\right) + \Omega^2_{\rm L} A_3 = F_{12} \dot{\phi}
\frac{c}{a^2} \,. \label{Maxw44}
\end{equation}
Here and below the dot denotes the derivative with respect to time. The corresponding equation for the axion field takes now the form:
\begin{gather}
\label{bas10b}
\ddot \phi+ \left(\frac{2\dot
a}{a}+\frac{\dot c}{c}\right) \dot \phi + m^2_{({\rm A })}\phi = {\dot{A}}_3 \ \frac{F_{12}}{\Psi^2_0 a^2 c}\,.
\end{gather}
We deal with a coupled system of equations, which displays three
interesting features. First, when  $F_{12}=0$ these two equations
decouple, and the appropriate solution for $A_3(t)$ with initial
value $A_3(t_0)=0$ is the trivial solution $A_3 \equiv 0$. In
other words, in the absence of magnetic field there are no reasons
for the longitudinal electric filed production. Second, when
$F_{12}\neq 0$, and $\dot{\phi} \neq 0$, the electric field is
inevitable, since the solution $A_3(t)=0$ is not admissible.
Third, when plasma is absent, (\ref{Maxw44}) gives
\begin{equation}
{\dot{A}}_3 = F_{12} \frac{c}{a^2} \ \phi(t)   \,, \label{Maxw441}
\end{equation}
(we put here $\phi(t_0)=0$ for simplicity), and the equation
(\ref{bas10b}) converts into
\begin{gather}
\label{bas10b99} \ddot \phi+ \left(\frac{2\dot a}{a}+\frac{\dot
c}{c}\right) \dot \phi+ \phi \left[m^2_{({\rm A })} -
\frac{F^2_{12}}{\Psi^2_0 a^4} \right] = 0 \,.
\end{gather}
As it was shown in \cite{BBT} this model describes anomalous
growth of the pseudoscalar field in the interval of the
cosmological time, when $m^2_{({\rm A })} <
\frac{F^2_{12}}{\Psi^2_0 a^4(t)}$. The formula (\ref{Maxw441})
explains one very important detail inherent in the model: under
the influence of the axion field pure magnetic field transforms
into the pair of parallel magnetic and electric fields. In the
paper \cite{BG} we indicated such field configurations as
longitudinal magneto-electric clusters. The presence of the term
$\Omega^2_{\rm L} A_3$ in (\ref{Maxw44}) changes the behavior of
the model principally. We expect now the appearance of solutions
of the oscillatory type, and the cooperative electromagnetic field
in the Vlasov plasma plays here the key role. Indeed, when the
axion field produces electric field from the magnetic one, this
electric field tries to separate plasma particles with positive
and negative charges. However, this procedure generates the
response of the cooperative electric field in plasma due to the
Vlasov mechanism. Since the cosmological evolution is a
non-stationary process, the interaction of axion-induced and
cooperative fields leads to oscillations in plasma. Let us
describe this process in more details.

\section{Dispersion relations}

First of all, let us discuss the hierarchy of time scales, which
appear in the model.

The first time scale is predetermined by the Universe expansion. When one deals with isotropic universe expansion,
effectively this time scale can be estimated using the Hubble function $\tau^{-1}_{\rm cosmo} \to {\cal H}(t){=}\frac{\dot{a}}{a}$. When we deal with anisotropic
cosmological model, we have to use the maximal rate of expansion $\tau^{-1}_{\rm cosmo} \to {\rm max \left\{ \frac{\dot{a}}{a}, \frac{\dot{c}}{c}\right\}}$.
At present the value of the parameter $\tau^{-1}_{\rm cosmo}$ is of the order $\tau^{-1}_{\rm cosmo} \simeq {\cal H}(t_0) \simeq 10^{-18}\ {\rm s}^{-1}$;
in the recombination epoch it was of the order $\tau^{-1}_{\rm cosmo} \to {\cal H}(t_R) \simeq 10^{-13}\ {\rm s}^{-1}$; in the early Universe, near the inflation epoch,
when the anisotropy could be essential, one can use the estimation $\tau^{-1}_{\rm cosmo} \to 10^{-2}\ {\rm s}^{-1}$.

The second time scale relates to the frequency associated with the reduced axion mass; clearly, the mass range
$10^{-6}\ {\rm eV} \lesssim m_{(\rm axion)}\lesssim 10^{-2}\ {\rm eV}$ corresponds to the frequency range $10^{9}\ {\rm s}^{-1}\lesssim m_{(\rm A)}\lesssim 10^{13}\ {\rm s}^{-1}$.

The third time scale corresponds to the Langmuir frequency; for instance, for the thermonuclear plasma the estimations yield
$\Omega_L \simeq 10^{11} \ {\rm s}^{-1}$.

Finally, the magnetic field $B$ and the axion-photon coupling constant $\frac{1}{\Psi_0}$ define the fourth time scale parameter  $\frac{B}{\Psi_0}$; for the range
$10^{5}\ {\rm GeV} \lesssim \Psi_0 \lesssim 10^{12}\ {\rm GeV}$ and the magnetic field $B \to 1 \ {\rm G}$ this parameters belongs to the interval
$10^{-6}\ {\rm s}^{-1}\lesssim \frac{B}{\Psi_0}\lesssim 1\ {\rm s}^{-1}$; for the relativistic rapidly rotating neutron stars  with $B \to 10^{12} \ {\rm G}$, the basic quantity
$\frac{B}{\Psi_0}$ is about $10^{12}$ times bigger; for the interstellar plasma with $B \to 10^{-6} \ {\rm G}$ it is, correspondingly, about $10^{6}$ times smaller.

In our consideration we assume that the quantity  $\tau^{-1}_{\rm cosmo}$, is much smaller than the quantities
$\Omega_{\rm L}$, $m_{({\rm A })}$, $\frac{\dot{\phi}}{\phi}$, and $\frac{B}{\Psi_0}$; for sure, one can find a lot of cosmological epochs, for which our assumption could be valid. In
other words, we assume that the process of plasma self-regulation
is much more quick than the process of the universe expansion.
Keeping in mind this restriction, we put below $a(t_0)$ and
$c(t_0)$ into the master equations of axion electrodynamics
instead of functions $a(t)$ and $c(t)$.

Then to analyze the model we can use the standard Fourier-Laplace
transformation
\begin{equation}
\mathcal{A}(\Omega)=\int\limits_{t_0}^{\infty} dt \, A_3(t)\, {\rm
e}^{i\, \Omega (t-t_0)} \,, \label{f1}
\end{equation}
\begin{equation}
\varphi(\Omega)=\int\limits_{t_0}^{\infty} dt \,
\phi(t)\, {\rm e}^{i\, \Omega (t-t_0)} \,. \label{f2}
\end{equation}
The corresponding equations for the Fourier-Laplace images take
the form
\begin{gather}
 \mathcal{A} \, [-\Omega^2+\Omega_{\rm L}^2]+ i\,\Omega \, \varphi \, F_{12} \frac{c(t_0)}{a^2(t_0)} = j_1 \,,
 \nonumber \\
i\,\Omega \, \mathcal{A} \, \frac{F_{12}}{\Psi^2_0 a^2(t_0)c(t_0)}
+\varphi \, [-\Omega^2+m^2_{({\rm A })}] = j_2\,, \label{f4}
\end{gather}
where
\begin{gather}
j_1 \equiv  \dot{A}_3(t_0) -i\, \Omega A_3(t_0) - F_{12}
\frac{c(t_0)}{a^2(t_0)} \phi(t_0)\,, \nonumber\\
j_2 \equiv  \dot{\phi}(t_0)-i\, \Omega \phi(t_0) - A_3(t_0) \
\frac{F_{12}}{\Psi^2_0 a^2(t_0)c(t_0)} \,. \label{f5}
\end{gather}
Clearly, all the information about poles of the Fourier-Laplace
images $\mathcal{A}(\Omega)$ and $\varphi(\Omega)$ is encoded in
the determinant of the system (\ref{f4})
\begin{gather}
\Delta = \Omega^4 - \Omega^2 \left(\Omega^2_{\rm L} + m^2_{({\rm
A})}-\frac{B^2}{\Psi_0^2}\right)+m^2_{({\rm A })}\Omega^2_{\rm L}\,,
\label{f6}
\end{gather}
where we introduced the constant $B^2=\frac{F_{12}^2}{a^4(t_0)}$
for the sake of simplicity. Let us discuss, first, three
interesting particular solutions of the dispersion equation
$\Delta(\Omega)=0$.

\subsection{Three special solutions}

\subsubsection{Magnetic field is absent, $F_{12}=0$}

In this case the dispersion equation gives two decoupled
solutions: first, $\Omega=\pm \Omega_{\rm L}$, which describes
pure plasma oscillations; second, $\Omega=\pm m_{({\rm A })}$,
which relates to the axion induced oscillations. The same result
can be obtained, when $\frac{1}{\Psi_0}=0$, i.e., the axion-photon
coupling constant is equal to zero.

\subsubsection{Massless axions, $m_{({\rm A })}=0$}

In this case one obtains that the first (double) root is $\Omega
=0$, and other two are the following: $\Omega {=} \pm
\sqrt{\Omega^2_{\rm L}{-} \Omega^2_{\rm B}}$, where $\Omega^2_{\rm
B} \equiv \frac{B^2}{\Psi_0^2}$. We deal with oscillations, when
$\Omega^2_{\rm L} > \Omega^2_{\rm B}$. When $\Omega^2_{\rm L}<
\Omega^2_{\rm B}$, i.e., the magnetic field is strong or/and the
axion-photon coupling constant $\frac{1}{\Psi_0}$ is rather big,
we obtain non-harmonic processes in plasma: one increasing mode
and one decreasing mode exist.

\subsubsection{Langmuir frequency is small}

When $\Omega^2_{\rm L} \to 0$, we obtain the double solutions
$\Omega=0$ and the solutions $\Omega=\pm \sqrt{m^2_{({\rm A
})}-\Omega^2_{\rm B}}$. Again, we deal with oscillations, when
$m^2_{({\rm A })}> \Omega^2_{\rm B}$, and with non-harmonic
process, when $m^2_{({\rm A })}<\Omega^2_{\rm B}$.

\subsection{General case:  $m_{({\rm A })}\neq 0$, $\Omega_L\neq 0$, $F_{12}\neq 0$, $\frac{1}{\Psi_0}\neq 0$}

In order to facilitate the analysis of this general situation, let
us introduce the the combination frequencies:
\begin{gather}\label{odr1}
\Omega_{\pm} \equiv \Omega_{\rm L} \pm m_{({\rm A })} \,.
\end{gather}
The solution of biquadratic equation $\Delta(\Omega)=0$ can be
written as
\begin{gather}\label{odr3}
2\Omega^2 = \frac{1}{2}(\Omega^2_+ + \Omega^2_-)-\Omega^2_{\rm B}
\pm \sqrt{(\Omega^2_+ -\Omega^2_B)(\Omega^2_- -\Omega^2_{\rm B})}.
\end{gather}
Let us classify these roots.

\subsubsection{Two different pairs of real roots}

There are two different branches of oscillations, when
\begin{gather}\label{odr7}\begin{cases}
&(\Omega^2_+ - \Omega^2_{\rm B})(\Omega^2_- - \Omega^2_{\rm B}) > 0 \,, \\
&\frac{1}{2}(\Omega^2_+ - \Omega^2_-) + (\Omega^2_- - \Omega^2_{\rm B}) > 0 \,.
\end{cases}
\end{gather}
Clearly, it is possible, when
$\Omega^2_{\rm B}<\Omega^2_{-}$, thus, the formula
\begin{gather}\label{odr966}
\Omega = \pm \frac12 \left[\sqrt{\Omega^2_{+} - \Omega^2_{\rm B}} \pm  \sqrt{\Omega^2_{-} - \Omega^2_{\rm B}} \right]
\end{gather}
is well-defined. As an illustration, let us calculate the electric
field $E(t)=-\dot{A}_3$, which relates to the solution
(\ref{odr966}) and to the initial data $\dot{A}_3(t_0)=0$,
$A_3(t_0)=0$, $\dot{\phi}(t_0)=0$:
\begin{gather}
    E(t)=E_0\sin \left[\frac12\sqrt{\Omega^2_+
-\Omega^2_B}\,(t-t_0)\right]\,\nonumber \\ {}\times \sin
\left[\frac12\sqrt{\Omega^2_- -\Omega^2_B}\,(t-t_0)\right],
\end{gather}
where
\begin{gather}\label{FL1}
E_0=  \frac{2\,\phi(t_0) \ m^2_{({\rm A })} F_{12} \
\frac{c(t_0)}{a^2(t_0)}}{\sqrt{(\Omega^2_+ -\Omega^2_B)(\Omega^2_-
-\Omega^2_{\rm B})}} \,.
\end{gather}
Thus, the maximum value of the total electric field in plasma is
proportional to the modulus of the starting value of the
pseudoscalar field $|\phi(t_0)|$, to the square of its mass
$m^2_{({\rm A })}$, and to the modulus of the initial value of the
magnetic field $|F_{12}|$.

\subsubsection{Two coinciding pairs of real roots}

The solution of this type is of the form
\begin{gather}\label{odr9669}
\Omega = \pm \sqrt{m_{({\rm A})} \Omega_L} \,,
\end{gather}
it is possible, when $\left|\frac{B}{\Psi_0}\right| =|\Omega_L -
m_{({\rm A })}|$. Since the corresponding pole in the function
${\cal A}(\Omega)$ is the double one the contribution of this pole
is linear in the cosmological time $t$; in other word this case
relates to the resonance-type growth of the electric field
oscillations.

\subsubsection{Two pairs of pure imaginary roots }

We deal with two different pairs of pure imaginary roots, when
$|\Omega_{\rm B}| > \Omega_{+}$. When  $\Omega_{\rm L} + m_{({\rm
A })} = \left|\frac{B}{\Psi_0}\right|$ the pairs coincide yielding
$\Omega = \pm i \sqrt{m_{({\rm A})}\Omega_{\rm L}}$. Keeping in
mind the Fourier-Laplace transformations (\ref{f1}), (\ref{f2}),
we see that one of the modes is the damping one, while the another
mode is increasing and is proportional to $e^{\sqrt{m_{({\rm
A})}\Omega_{\rm L}} \ t}$.

\subsubsection{Complex roots: $\Omega = \alpha + i\gamma $}

This is the most wide class of solutions, it relates to the following requirements:
\begin{gather}\label{odr19}
\Omega^2_{-} < \Omega^2_{\rm B} < \Omega^2_{+} \,,
\end{gather}
and describes two modes of quasi-oscillations: increasing mode and
decreasing mode.

\section{Discussion}\label{sec6}

In the framework of the Maxwell-Vlasov-axion model we described
oscillations of electric field in an axion-active relativistic
plasma, which are assumed to be produced in the course of
anisotropic expansion of early universe with global magnetic
field. Let us shortly summarize the results of this analysis.

\noindent 1. The physical mechanism of the cosmic electric field
oscillations seems to be the following. During the anisotropic
stage of expansion of the early universe a specific phase
transition is predicted to take place, which produces strong
magnetic field (see, e.g., the review \cite{TurnerR}). The
pseudoscalar (axion) field produced in early universe by the
Peccei-Quinn mechanism \cite{Peccei} transforms this (pure)
magnetic field into the magneto-electric field, the electric
component being parallel to the magnetic one and proportional to
the pseudoscalar field. Such magneto-electric configuration
(indicated as a longitudinal cluster in \cite{BG}), in its turn,
provokes the exponential growth of the axion number and then, we
obtain again the anomalous growth of the electric field. This
inflationary-type procedure can be stopped due to the gravity
field evolution, when the isotropic phase of the universe
expansion replaces the anisotropic one (see \cite{BBT}). As we
have shown in this paper, the growth of the axion number and of
the electric field, respectively, can be stopped by a more simple
mechanism, namely, by the cooperative the Vlasov field
counteraction in the relativistic plasma. Indeed, the axionically
induced electric field (this field has no electrically charged
sources, it is generated due to the axion-photon interaction in
the strong magnetic field) separates particles with negative and
positive charges thus generating the cooperative electric field,
which counteracts to the global electric field. Since the universe
expands, the described process is non-stationary, and the
generated cooperative electric field tends to be in an
instantaneous balance with global axionically induced electric
field.

\noindent 2. The growth of the total electric field can be
stopped, when, first, the electrodynamic system is in an pure
oscillatory regime; second, when it decreases with time
non-harmonically; third, when a quasi-periodic oscillations with
damping take place. In order to clarify the possibilities of the
mechanism described above, we have analyzed the dispersion
relations for the evolution of the coupled pseudoscalar (axion)
and electromagnetic fields. We have found that every damping mode
of quasi-oscillations is accompanied by the corresponding
increasing mode, thus, the only one interesting case exists,
namely, the pure oscillatory regime, for which the growth of the
axion number and of the electric field happened to be stopped by
the cooperative Vlasov field. These situations can be realized,
when $\Omega^2_{\rm B}<\Omega^2_{-}$; two frequencies of such
oscillations are of the form (\ref{odr966}); the estimation of the
maximum value of the electric field is given by the formula
(\ref{FL1}).

\noindent 3. We indicated the studied plasma model as axion-active
plasma, since its internal state is predetermined by the state of
global electric field produced by the pseudoscalar (axion) field
in the presence of non-stationary magnetic field. This term is
considered to be an analog of the term magneto-active plasma. In
the nearest future we hope to consider the dispersion relations
for the {\it waves} in the axion-magneto-active plasma by using
the solutions obtained here as a zero-order approximation in the
problem of kinetics of axion-electromagnetic perturbations in the
relativistic plasma.

\appendix

\acknowledgments The work was partially supported by the Russian
Foundation for Basic Research (Grant No. 14-02-00598).

\end{document}